\documentclass{article}
\usepackage{mrs2005,epsfig}
\def\fp{\hbox{f}_{\rm p}}
\def\ftype{\hbox{f}_{\rm type}}
\def\Bj{$b_{\rm J}$}
\def\MBj{$M_{b_{\rm J}}$}
\def\Halpha{$\rm H\alpha$}
\def\svi {\sigma(v)}
\def\gsim{\mathrel{\raise0.35ex\hbox{$\scriptstyle >$}\kern-0.6em 
\lower0.40ex\hbox{{$\scriptstyle \sim$}}}}
\def\lsim{\mathrel{\raise0.35ex\hbox{$\scriptstyle <$}\kern-0.6em 
\lower0.40ex\hbox{{$\scriptstyle \sim$}}}}
\def\fS0{\hbox{f}_{\rm S0}}

\def\kmsMpc{$\rm kms^{-1}Mpc^{-1}$}

\begin{document} 
\title{The CNOC2 sample of intermediate redshift galaxy groups - the powerhouse of galaxy evolution}

\author{D.~J.~Wilman$^{1}$,~M.~L.~Balogh$^{2}$,~R.~G.~Bower$^3$,~J.~S.~Mulchaey$^4$,~A.~Oemler~Jr$^4$,~R.~G.~Carlberg$^5$}
\affil{
$^1$Max-Planck-Institut f\"ur extraterrestrische Physik, Giessenbachstra\ss e, D-85748 Garching, Germany\\
$^2$Department of Physics, University of Waterloo, Waterloo, Ontario, Canada N2L 3G1\\
$^3$Physics Department, University of Durham, South Road, Durham DH1 3LE, U.K.\\
$^4$Observatories of the Carnegie Institution, 813 Santa Barbara Street, Pasadena, California, U.S.A.\\
$^5$Department of Astronomy, University of Toronto, Toronto, ON, M5S 3H8 Canada.\\
} 
 
\begin{abstract} 
The evolution of galaxies in groups may have important implications for the global evolution of star formation rate in the Universe, since many processes which operate in groups may suppress star formation, and the fraction of galaxies bound in groups at the present day is as high as $\sim$60\%. We present an analysis of our sample of 0.3$\leq$z$\leq$0.55 groups, selected from the CNOC2 redshift survey and supplemented with deep spectroscopy and HST ACS imaging.
We find that these groups contain significantly more passive galaxies than the field, with excesses of S0, elliptical and passive spiral galaxy types. The morphological composition is closely matched to that of more massive irregular clusters at a similar epoch. Contrasting with galaxy samples in a variety of environments and epochs, we find that the fraction of passive galaxies (EW[OII]$<$5A), is strongly evolving in the group environment, with parallel evolution in the (global) field population, whilst little evolution is observed in cluster cores since $z\sim 1$.
\end{abstract} 
 
\vspace*{-0.75cm}
\section{Introduction} 
 
Groups of galaxies represent the most common galaxy environment 
at $z\sim 0$, containing as much as 60\% of the galaxy population at
the present day \cite{Eke04}. 
Due to their relatively low velocity dispersions, groups of galaxies
are ideal sites for galaxy-galaxy interactions  
which are likely
to give rise to significant changes in the SFRs, total stellar masses,
and morphological appearance of galaxies. 
Recent studies in the local Universe show that galaxy properties are 
indeed influenced by the group environment (e.g. \cite{Balogh04}). 
Therefore, understanding the evolutionary processes in groups will probe 
not only the dependence of galaxy properties on environment, but also 
trace the importance of this common environment in driving global trends 
such as the strong decline in volume--averaged star formation since $z\sim 1$ 
(e.g. \cite{Madau98}).

Tracing galaxy evolution in groups requires the existence of higher redshift 
group catalogues. 
Distant groups have always been difficult to recognize because of their
sparse galaxy populations. The presence of high redshift groups has
typically been inferred indirectly via the presence of a radio galaxy
or X-ray emitting intragroup medium 
\cite{AllSmith93,Jones02b}. 
However, these selection criteria detect only the 
richest, elliptical dominated groups.
To reconcile this selection bias, 
large spectroscopic field surveys can be used to study galaxy groups selected 
purely on the basis of their three dimensional galaxy density. 
Such catalogues of groups now exist.

\section{The CNOC2 group sample at $0.3 \leq z \leq 0.55$}

The second Canadian Network for Observational Cosmology 
Redshift Survey (CNOC2) consists of $UBVR_{C}I_{C}$ photometry 
and spectroscopy, measuring $\sim $ 10$^4$ galaxy redshifts up to $z \sim 0.6$ \cite{Yee00}.
This provided an powerful opportunity to generate a kinematically selected sample of 
intermediate redshift galaxy groups. 
A friends-of-friends percolation algorithm was used to detect
groups of galaxies in redshift space \cite{Carlberg01}. 

This sample can be used to 
study the influence of the typical group 
environment on galaxy properties 
at intermediate redshift. 
To this end, we have recently obtained deep {\it Hubble Space Telescope} ACS imaging 
of a sub-sample of 26\footnote{The 26 groups include 20 targetted groups and 6 serendipitous groups.} 
CNOC2 groups 
with 
redshifts $0.3<z<0.55$, when the Universe was 
$\sim \frac{2}{3}$ its present age. 
The original spectroscopy has been supplemented with deeper
spectroscopy using the {\it LDSS-2} spectrograph on the 
{\it Magellan 6.5-m}
telescope \cite{Wilman05a} and will be further supplemented by deep 
{\it VLT FORS2} spectroscopy (almost complete at the time of writing), 
reaching R$_C \sim 23.2$ ($M_{*} + 3$ at $z = 0.4$).
Combined with forthcoming/ongoing observations at UV ({\it GALEX}), Near
Infra-red ({\it SOFI} on the {\it ESO NTT}) and X-ray ({\it Chandra}) wavelengths, 
we are building 
a unique catalogue of galaxy groups, combining significant look--back time with the 
depth 
required to probe the faint-end of the galaxy luminosity function, and coverage 
across the 
electomagnetic spectrum.

Magellan observations have extended the depth of our spectroscopic sample to 
R$_C$ = 22.0, down to which we have well--understood selection, and no bias towards 
emission--line galaxies \cite{Wilman05a}. Our combined group galaxy sample numbers 
282 galaxies in 26 separate groups at $0.3<z<0.55$.\footnote{For further details of the 
Magellan spectroscopy and group membership allocation, see Wilman~et~al. (2005a) 
\cite{Wilman05a}}
Within the area of (deeper) Magellan 
coverage, our spectroscopic sample also contains 334 serendipitous field galaxies within this 
redshift range. 

Throughout this paper we assume a $\Lambda$CDM cosmology of
$\Omega_{M} = 0.3$, $\Omega_{\Lambda}=0.7$ and $H_0 = 75$\kmsMpc.

\section{Results and Analysis}
The observed magnitudes of all galaxies in our sample are extinction corrected based 
upon the models of 
Schlegel~et~al. (1998) \cite{Schlegel98}. A best-fit SED is computed using 
observed local SEDs \cite{KE85} as templates, and the observed $R_c$-magnitude, 
(B-$I_c$) colour and spectroscopic redshift of each galaxy. 
The rest--frame \Bj\ luminosity (\MBj) is computed 
from the best-fit SED. This band is chosen because it 
closely matches 
the CNOC2 
spectroscopic selection band ($R_c$) 
in the observed--frame, and it facilitates 
comparisons with the 2PIGG local group catalogue.

\subsection{Star Formation and Evolution in CNOC2 Groups}

To study the relative levels of star formation in statistical galaxy
samples, we use the [OII]$\lambda3727$ emission line which lies
centrally in the visible window at $0.3\leq
z \leq 0.55$ and at wavelengths of low sky emission. 
We note that due to uncertainties in the calibration of star formation rates,
we deliberately limit our study to direct comparisons between measurements of the 
Equivalent Width (EW) of [OII]. Normalisation by the continuum also reduces 
uncertainties related to absorption by dust and aperture bias.

We are motivated by the clear bimodal distribution in colour and EW[\Halpha] 
(e.g. \cite{Balogh04,BaloghBaldry04}) which show that the fraction of red, 
passive galaxies is strongly dependent upon local galaxy density.
To examine how the fraction of passive galaxies depends upon environment in 
our sample, we impose 
an arbitrary division at EW[OII]=5\AA. 
We expect the population with EW[OII]$<$5\AA\ to be dominated by passive
galaxies and the
population with EW[OII]$\geq$5\AA\ to be dominated by star-forming
galaxies. This division is sufficient to reveal trends in
statistical samples (e.g. \cite{Hammer97}).
To assess the relative size of each population, we define {\bf$\fp$} to equal the 
weighted\footnote{Galaxy weights are computed to account for incompleteness in 
the sample. See 
Wilman~et~al. (2005a) \cite{Wilman05a} for a derivation of selection functions and 
weights in this sample.} 
fraction of galaxies with EW[OII]$<$5\AA\ (i.e. an estimator of the fraction of {\it passive} 
galaxies). 

\begin{figure}  
\vspace*{1.25cm}  
\begin{center}
\centerline{\epsfig{figure=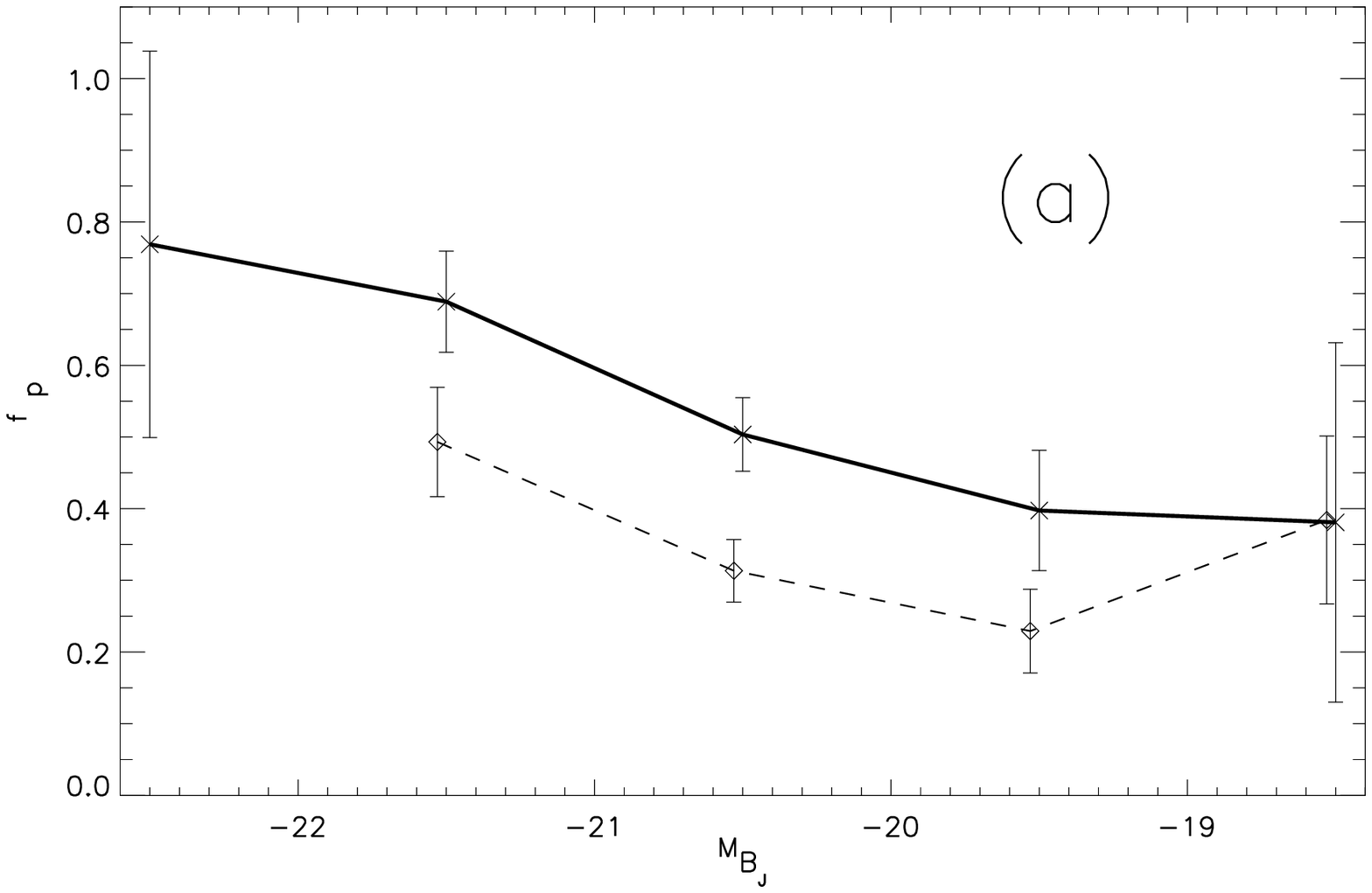,width=0.47\textwidth,height=5cm}\epsfig{figure=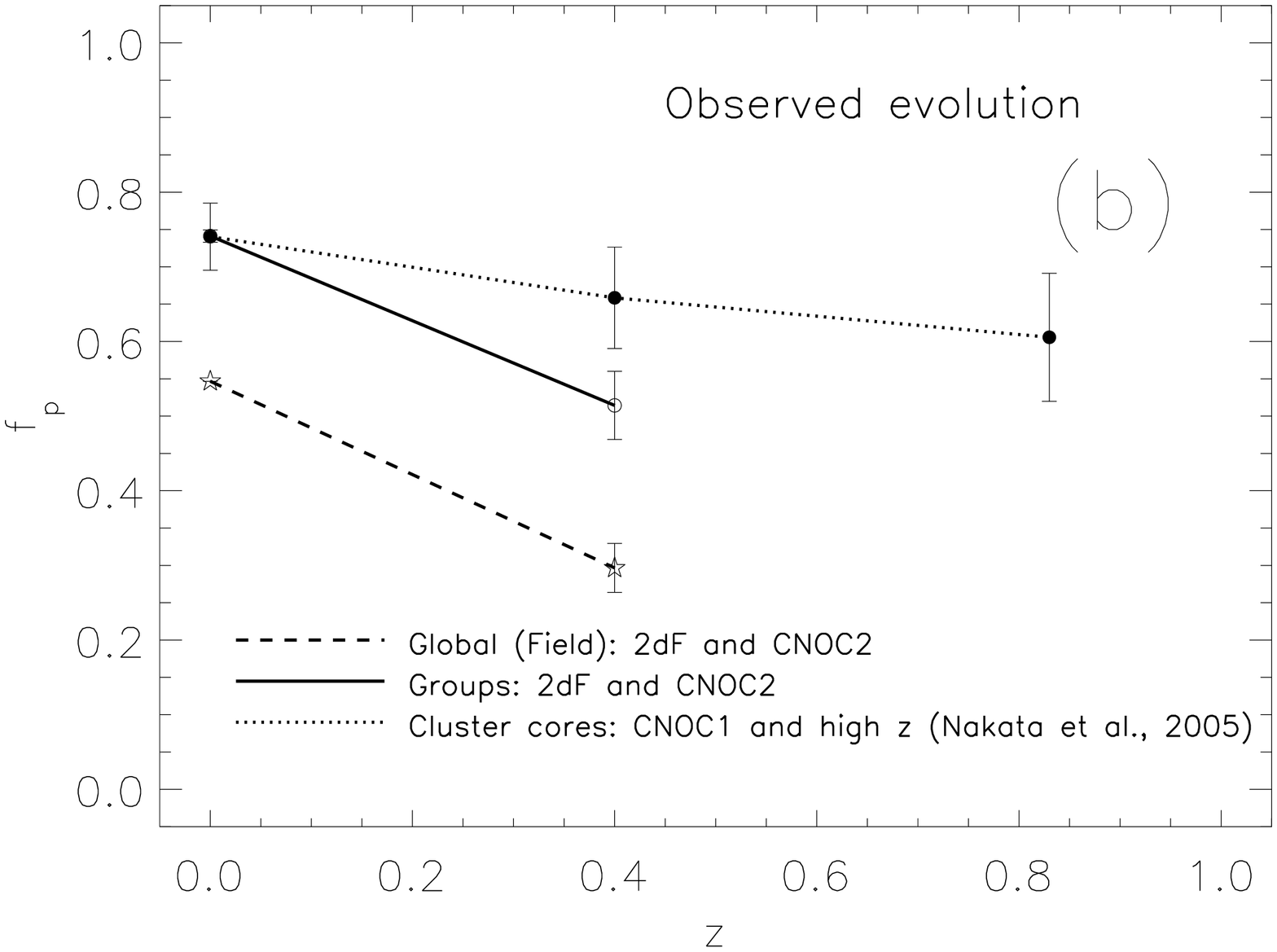,angle=90,width=0.47\textwidth,height=5cm}}  
\end{center}
\vspace*{-0.25cm} 
\caption{{\bf (a)} The fraction of passive galaxies, $\fp$, in the
stacked group within $1h_{75}^{-1}$Mpc of the 
group centre (solid line) and the field (dashed line) as a function of
galaxy luminosity, \MBj. The field symbols are offset slightly in
luminosity for clarity. Statistical errors on $\fp$ are computed using
a Jackknife method.
{\bf (b)} The fraction of passive galaxies, $\fp$, plotted as a function of redshift in different environments. The group and field data are selected from our 2dF and CNOC2 samples. We overplot the equivalent fractions in cluster cores, selected from the samples of \cite{Nakata05}. These data originate in the 2dFGRS, CNOC1 and high redshift cluster samples of \cite{vanDokkum00} and \cite{Postman01}. Cluster data are selected only within the inner $0.75h_{75}^{-1}$~Mpc of the cluster centre and we match the luminosity limit of \cite{Nakata05} (\MBj$=-19.65$). This figure shows the strong decline of $\fp$ in the Universe since $z \sim 0.4$ in groups and the field whilst the passive population in the dense core regions of clusters is consistent with having been mostly in place since $z \sim 1$.
} 
\label{fig:fplum}
\end{figure}

Figure~\ref{fig:fplum}{\bf (a)} shows how 
$\fp$, depends upon galaxy \Bj-band luminosity in the group and
field samples. The group galaxy sample is limited to those within $1h_{75}^{-1}$Mpc 
of the 
luminosity-weighted group centroid. 
There is a clear enhancement of
$\fp$ in the group galaxies with respect to the field, especially in
the luminosity range $-22.0 \leq $\MBj$\leq -19.0$. Combining all galaxies within the 
luminosity range $-22.5 \leq$\MBj$\leq -18.5$, the enhancement in groups of $\fp$ is 
better than $3\sigma$ significance, and this trend is still evident if the
most massive two groups (velocity dispersion $\svi > 600$~km~s$^{-1}$) 
are excluded from the sample.

We interpret the enhancement of $\fp$ in groups as direct evidence that galaxies in 
intermediate redshift groups are significantly less likely to have ongoing star 
formation than field galaxies. This could be a related to either a different 
formation history
in the group environment, intragroup environmental processes
accelerating galaxy evolution to the passive state, or a combination
of these two effects.
We also note that their is a strong luminosity dependence on $\fp$ in both group and 
field samples, such that a higher fraction of brighter galaxies are passive by 
z$\sim$0.4. This is consistent with the overall picture that more massive galaxies 
have formed all their stars at earlier epochs.

%

In Figure~\ref{fig:fplum}{\bf (b)} we examine evolution in the cluster, group and field 
environments. For our low redshift sample, we choose the 2dFGRS field survey and the 
2PIGG group catalogue \cite{Eke04}. A detailed description of sample selection, and the 
comparison between CNOC2 and 2dFGRS data 
 can be found in Wilman et al. (2005b) \cite{Wilman05b}.
The cluster core data are taken from Nakata et al. (2005) 
\cite{Nakata05}\footnote{Note that whilst in that 
paper a 10\AA\ division in EW[OII] is used, we maintain consistent use of a 5\AA\ limit.}
, originating in the 2dFGRS, CNOC1 and high redshift cluster samples of 
Van Dokkum et al. (2000) \cite{vanDokkum00} 
and Postman et al. (2001) \cite{Postman01}. 
These data are limited to within $0.75h_{75}^{-1}$~Mpc of the cluster centres. 
To match the Nakata et al. (2005) \cite{Nakata05} sample we apply a luminosity limit of \MBj$=-19.65$ to all data. This figure neatly illustrates the strength of evolution observed for $\fp$ in groups and the field whilst galaxies in dense cluster cores were already largely passive by $z \sim 1$. 

\subsection{Morphological Composition of CNOC2 Groups} 

With deep HST ACS (Hubble Space Telescope - Advanced Camera for Surveys) 
F775W filter observations of our CNOC2 groups we have 
the resolving power to classify galaxy morphologies to 
well below the magnitude limit of our spectroscopy. 
In this way we can isolate the morphological transformations which 
may be driving galaxy evolution in groups at intermediate redshift.

To date, a sub-sample of galaxies from the first 16 ACS fields have been 
morphologically classified visually by Augustus Oemler Jnr. (AO).
\footnote{We acknowledge that the analysis presented here is merely preliminary, as the 
classifications will be extended to the full sample, and classifications by more than 
one author will be used to assess the robustness of the classification process.}
Galaxies with spectroscopic redshifts in the $0.3<z<0.55$ redshift range 
were visually classified according to the MORPHS classification scheme \cite{Smail97}.
The current sample of morphologically classified CNOC2 galaxies includes 
158 group and 124 field galaxies.

\begin{figure}  
\vspace*{1.25cm}  
\begin{center}
\epsfig{figure=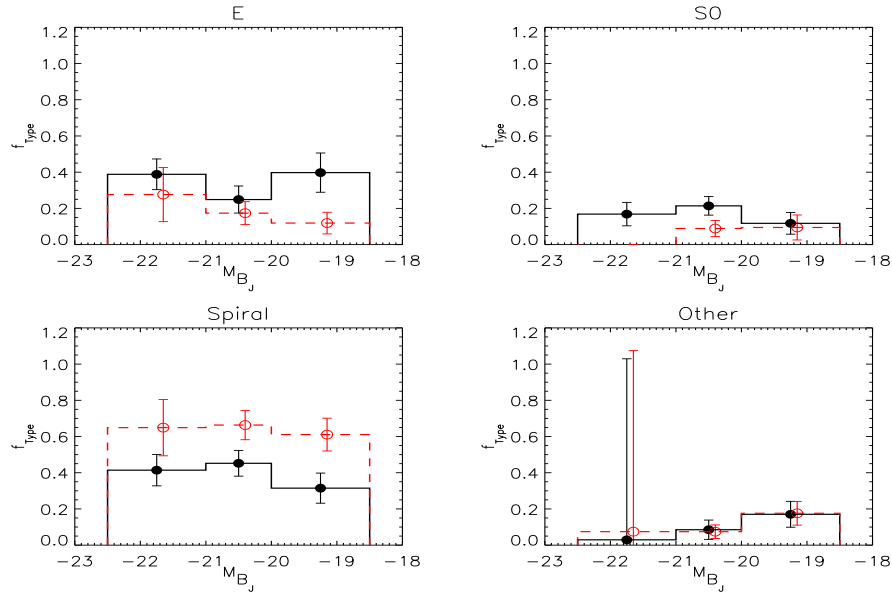,width=12.5cm,height=8.0cm}  
\end{center}
\vspace*{-0.5cm}
\caption{The fraction of each morphological type of galaxy, $\ftype$, in 3 bins of luminosity: Bright (\MBj$< -21.0$), Control ($-21.0\leq$\MBj$< -20.0$) and Faint ($-20.0\leq$\MBj$\leq -18.5$). The solid line with the filled circles represents the group galaxy population and the dashed line with the open circles represents the field population. Errors are estimated using the Jackknife technique and field points are offset slightly in \MBj\ for clarity. This figure clearly shows that the CNOC2 groups possess a larger fraction of early galaxy types (elliptical and S0) than the field, in place of spiral galaxies, more common in the field.} 
\label{fig:FtypeLum}
\end{figure}

To assess the statistical significance of differences between the morphological 
composition of group and field samples, we bin the galaxies into four broad 
morphological classes. These are Elliptical, S0, Spiral and Other (mostly Irregular 
and merger types).
In Figure~\ref{fig:FtypeLum} we show the fraction of galaxies of each morphological type, $\ftype$ 
in 3 bins of luminosity (\MBj$< -21.0$, $-21.0\leq$\MBj$< -20.0$, $-20.0\leq$\MBj$\leq -18.5$).\footnote{
For reference, in the local 2dFGRS survey, $M_{*} =-20.28$ in our cosmology.} 
The solid line with the filled circles represents the group galaxy population and the dashed line 
with the open circles represents the field population. 
It shows that whilst spiral galaxies dominate the field galaxy population, there is a significant deficit of spirals in galaxy groups ($3\sigma$ computed for all galaxies \MBj$\leq -18.5$). In the place of these spirals, there is an excess of early-type galaxies in groups. 
There is a $2.7\sigma$ excess of ellipticals and a $1.3\sigma$ excess of S0 type galaxies in total. Figure~\ref{fig:FtypeLum} indicates that the elliptical excess is most significant at faint luminosities, where the field elliptical population is less important. However at brighter luminosities, the S0 population seems to become very important in groups. In fact, brighter 
than \MBj$\lsim -20.5$ there are 16 group and no field S0 type galaxies. 
We compute the fraction of S0 galaxies in this luminosity range to be $\fS0 = 0.23$ in the groups. A 
bootstrap-resampling technique is then used to estimate the significance of this excess with respect to 
$\fS0 = 0$ in the field. 
We find that there is  
a $\sim 3.5\sigma$ excess of bright S0s in groups.
Finally, we also note that bright S0s appear to be particularly predominant in groups 
with velocity dispersions of $\sim 400$~km~s$^{-1}$.
Indeed, six of these galaxies are located in just two groups  
(3 in group 37, $\svi = 419\pm97$~km~s$^{-1}$ and 3 in group 39, 
$\svi = 454\pm88$~km~s$^{-1}$), 
suggesting that the formation of bright 
S0 galaxies appears to be biased towards a certain type of group environment, 
as represented by these groups.

A comparison with the $z \sim 0.5$ MORPHS cluster sample of Dressler et. al. (1997) 
\cite{Dressler97} indicates that our groups match the morphological composition of the 
irregular MORPHS clusters ($\sim 40\%$ spirals, $\sim 40\%$ ellipticals and $\sim 20\%$ 
S0s). This suggests that the morphological mix of unvirialised clusters is likely 
to be already in place during the infalling group stage.

%
%

\begin{figure}
\centerline{\epsfig{figure=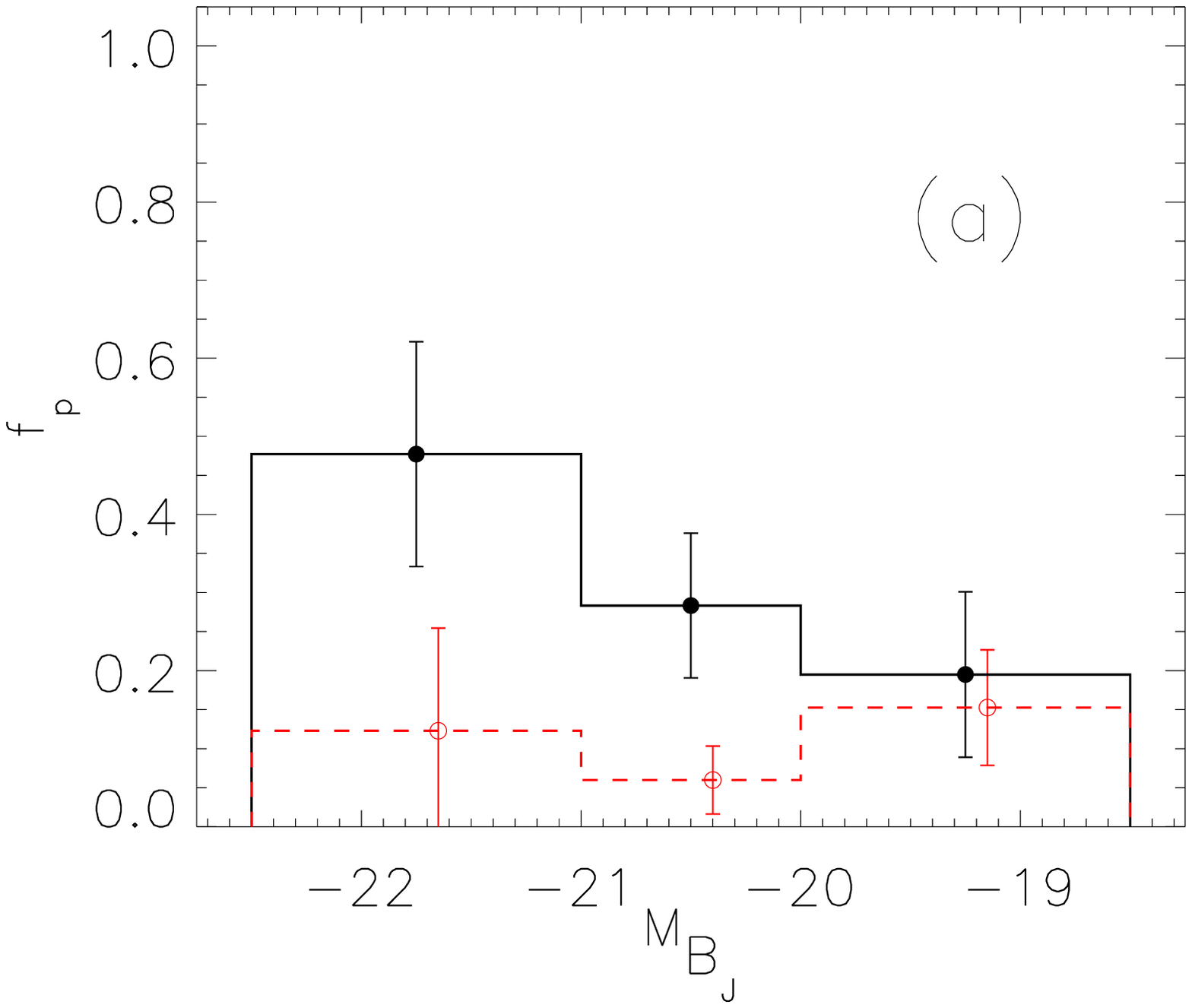,width=0.47\textwidth,height=5cm}\epsfig{figure=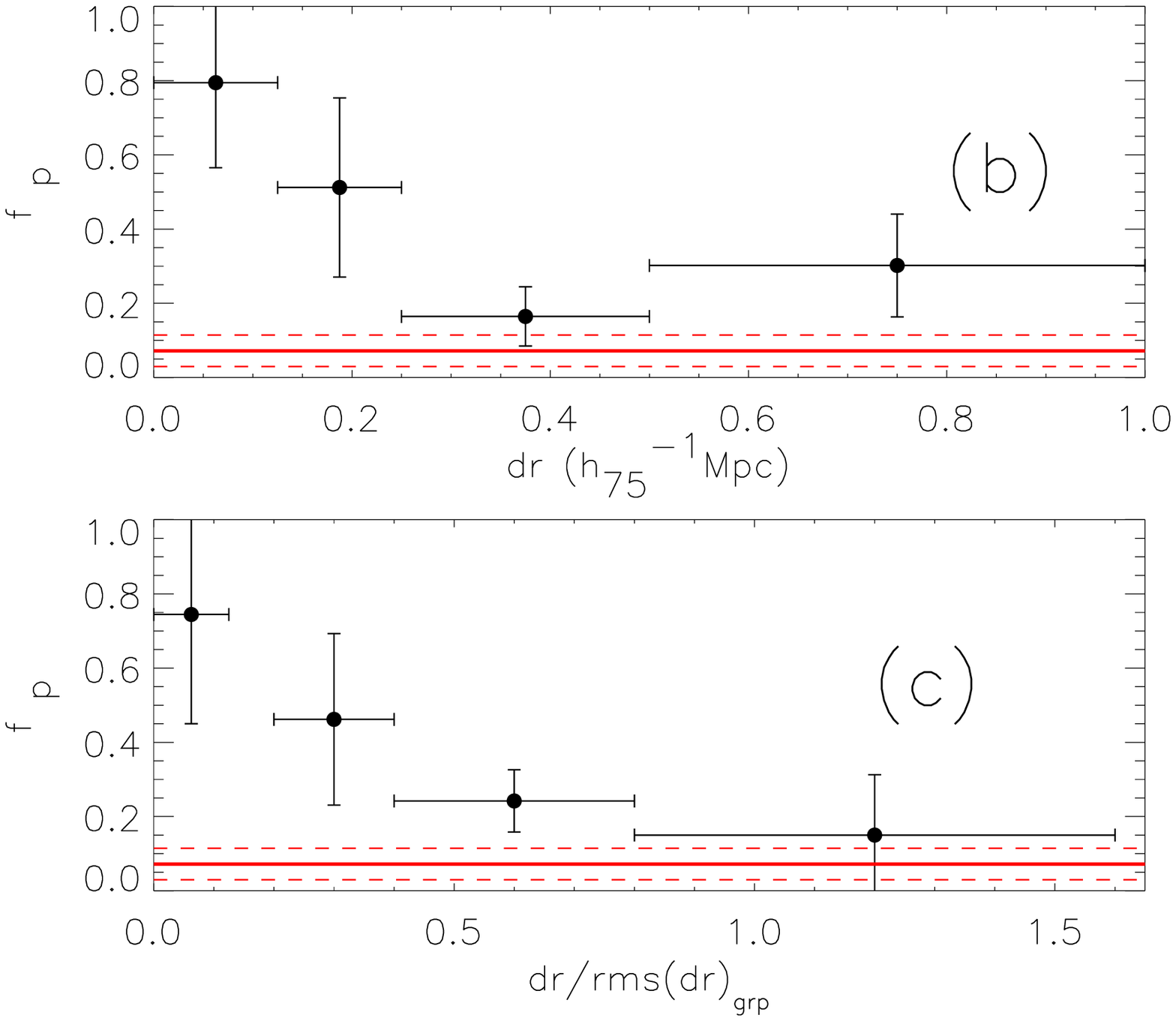,width=0.47\textwidth,height=5cm}}
\vspace*{-0.25cm} 
\caption{{\bf (a)} The fraction of spiral galaxies which are passive, $\fp$, in groups (solid line, filled circles) and in the field (dashed line, open circles) in 3 bins of luminosity: Bright (\MBj$< -21.0$), Control ($-21.0\leq$\MBj$< -20.0$) and Faint ($-20.0\leq$\MBj$\leq -18.5$). Errors are estimated using the Jackknife technique. The passive fraction in group spirals is only significantly greater than in the field brighter than \MBj$= -20.0$.
{\bf (b)} The fraction of spiral galaxies brighter than \MBj$=-20.0$ which are passive, $\fp$, in the field (solid horizontal line) and in groups as a function of group-centric radius, dr, and {\bf (c)} group-centric radius, normalised by the rms value of the parent group. The liklihood that a spiral galaxy is no longer forming significant numbers of stars is a clear function of its position in the group.}
\label{fig:fpLumSps}
\end{figure}

Another interesting class of galaxy is the Passive Spiral. These galaxies are still 
morphologically spiral-type, but with no sign of ongoing star formation (in our case, 
EW[OII]$<$5\AA). Passive spirals have typically aroused interest in cluster studies, 
where they have often been interpreted to be infalling spiral galaxies in which 
star formation has been somehow truncated without any morphological transformation, 
up to the point of observation (e.g. \cite{Couch98,Poggianti99}).
Figure~\ref{fig:fpLumSps}{\bf(a)} shows $\fp$ as a function of luminosity, only for 
galaxies which are morphologically classified spirals (67 in groups, 68 in the field).
At brighter luminosities (\MBj$\leq -20.0$), the population of spiral galaxies is 
clearly more passive in groups (solid line, filled circles) than the field 
(dashed line, open circles). 
This excess is assessed at $3.2\sigma$ signficance using a resampling method.
Interestingly, the lack of any passive spiral excess in fainter galaxies 
suggests that interactions with the IGM are unlikely to be responsible, since 
less massive galaxies have smaller gravitational potential wells, and thus should 
lose their gas more easily.
In the right-hand panels of Figure~\ref{fig:fpLumSps}, we examine whether $\fp$ in 
bright (\MBj$\leq -20.0$) spiral galaxies has any dependence upon its 
group-centric radius, dr {\bf(b)} and group-centric radius, 
normalised by the rms(dr) for the parent group {\bf(c)}. 
It is clear that a larger fraction of spiral galaxies are passive 
towards the group centre, and that this relation holds when the radial distance 
is normalised by the typical group-centric radius of each group.
This indicates that processes leading to the ceasation of star formation in bright spiral galaxies are tightly related to the group environment.
However, we note that we find no passive spiral in the two most massive groups 
($\svi >$600~km~s$^{-1}$). 
Since passive galaxies in local clusters seem to be found predominantly in the 
cluster outskirts 
\cite{GotoPassSp03}, we believe passive spirals are therefore most likely to be found in 
environments typical of smaller groups and the outskirts of clusters. 

\vspace*{-0.5cm}
\section{Conclusions} 

We have presented an analysis of galaxy properties in the group and field environments at intermediate redshift 
($0.3 \leq z \leq 0.55$). This is part of an ongoing study of how the group environment influences galaxy 
evolution. Our main conclusions to date can be summarized as follows:
 
\begin{itemize}

\item{Whist (massive) galaxies were largely passive in cluster cores by $z\sim 1$, there has been strong 
evolution in the fraction of passive galaxies ($\fp$) in groups and the field since $z\sim 0.4$. This must be 
important when considering the global budget of star formation.}

\item{Galaxies in CNOC2 groups are more likely to be passive or bulge-dominated than in the CNOC2 field.}

\item{However, the morphological composition of the z$\sim 0.4$ group population matches that of the irregular 
MORPHS clusters at similar redshift.}

\item{Excesses of bright S0s and passive spirals in CNOC2 groups relates the 2 galaxy types and suggests that the 
S0 formation process may favour groups. These excesses are not found fainter than $\sim M_*$ and so interactions 
with the IGM are unlikely to be responsible.}

\end{itemize}

\vspace*{-0.5cm}
\acknowledgements{ 

We acknowledge the CNOC2 survey team for providing an excellent dataset without which this study would not be 
possible. We are also grateful to those involved in the other studies from which we have compiled 
Figure~\ref{fig:fplum}{\bf (b)}.

}

\vfill 
\end{document}